\documentclass[pra,eqsecnum,twocolumn,amsfonts,superscriptaddress]{revtex4}
\usepackage{amsfonts}
\usepackage{amsmath}
\usepackage{bm}
\usepackage{epsf}

\newcommand{\beq}{\begin{equation}}
\newcommand{\eeq}{\end{equation}}
\newcommand{\beqa}{\begin{eqnarray}}
\newcommand{\eeqa}{\end{eqnarray}}

\newcommand{\ket}[1]{| #1    \rangle }
\newcommand{\bra}[1]{ \langle   #1  | }
\newcommand{\ave}[1]{  \langle #1   \rangle }
\newcommand{\qave}[2]{  \langle #1  | #2  | #1  \rangle }
\newcommand{\mel}[3]{  \langle #1  | #2   | #3  \rangle }
\newcommand{\amp }[2]{ \langle #1 |  #2  \rangle }

\newcommand{\weakv}[3]{\frac{ \mel{#1}{#2}{#3} }{ \amp{#1}{#3}} }

\newcommand{\vect}[1]{{\bm{ #1}}}

\newcommand{\rref}[1]{~(\ref{#1})}
\newcommand{\ccite}[1]{~\cite{#1}}

\newcommand{\wva}[1]{\mathcal{A}_{_{#1}} }

\newcommand{\post}{{{}_*}}

\begin{document}

%%%%%%%%%%%%%%%%%%%%%%%%%%%%%%%%%%%%%%%%%%%%%%%%%%%%%%%%%%%%%%%%%%%%%%%%
%%%%%%%%%%%%%%%%%%%% Local Definitions %%%%%%%%%%%%%%%%%%%%%%%%%%%%%%%%%

\newcommand{\diracsl}[1]{\not\hspace{-3.0pt}#1}

\newcommand{\pin}{\psi_{\!{}_1}}
\newcommand{\psf}{\psi_{\!{}_2}}
\newcommand{\psv}[1]{\psi_{\!{}_#1}}
\newcommand{\lab}[1]{{}^{(#1)}}
\newcommand{\psub}[1]{{\cal P}_{{}_{\! \! #1}}}
\newcommand{\sst}[1]{{\scriptstyle #1}}
\newcommand{\ssst}[1]{{\scriptscriptstyle #1}}
\newcommand{\aft}{{{\scriptscriptstyle\succ}}}
\newcommand{\bef}{{{\scriptscriptstyle\prec}}}

%%%%%%%%%%%%%%%%%%%%%%%%%%%%%%%%%%%%%%%%%%%%%%%%%%%%%%%%%%%%%%%%%%%%%%%%
%%%%%%%%%%%%%%%%%%%%%%%%%%%%%%%%%%%%%%%%%%%%%%%%%%%%%%%%%%%%%%%%%%%%%%%%

\title[Quantum Averages of Weak Values]
{Quantum Averages of Weak Values}

\author{ Yakir Aharonov  }
\affiliation{ Department of Physics and Astronomy, University of
South Carolina, Columbia, SC 29208} \affiliation{Department of
Physics and Astronomy, Tel Aviv University, Tel Aviv 69978,
Israel} \affiliation{Department of Physics and Astronomy, George
Mason University,  Fairfax, VA 22030 }
\author{  Alonso Botero }
\email{abotero@uniandes.edu.co} \affiliation{ Department of
Physics and Astronomy, University of South Carolina, Columbia, SC
29208}
\affiliation{
    Departamento de F\'{\i}sica,
    Universidad de Los Andes,
    Apartado A\'ereo 4976,
    Bogot\'a, Colombia }

\date{\today}

\begin{abstract}
\bigskip
We re-examine the status of the weak value of a quantum mechanical
observable as an objective physical concept, addressing its
physical interpretation and  general domain of applicability. We
show that the weak value can be regarded as a \emph{definite}
mechanical effect on a measuring probe specifically designed to
minimize the back-reaction on the measured system. We then present
a new framework  for  general measurement conditions (where the
back-reaction on the system may not be negligible)  in which  the
measurement outcomes can still be interpreted as \emph{quantum
averages of weak values}. We show that in the classical limit,
there is a direct correspondence between quantum averages of weak
values and posterior expectation values of classical dynamical
properties according to the classical inference framework.

\end{abstract}

\maketitle

\section{Introduction}

%In the macroscopic domain, measurements can be devised so
% that with negligible disturbance to the system under
% observation,   values of  dynamical quantities of
% interest can be ascertained with arbitrary precision.
% These conditions allow for an
%  effective description  of macroscopic objects in
%  terms of physical properties
% that may be regarded as  objective
% properties of the system, independently of the act of observation.
%In the quantum domain, however, a well-established
%trade-off\ccite{FuchsPeres}  between the information gain of a
%measurement and the accompanying disturbance to the measured
%system forbids the existence of passive measurements yielding
%arbitrarily sharp results. It has therefore been the general
%consensus that a description  in terms of objective physical
%properties of quantum systems, no description is possible in terms
%of independent of the .

In previous publications\ccite{AV90,AV91,RA95,Vaid96,Vaidman96b},
an objective description of a quantum system in the time interval
between two complete measurements has been proposed  in terms of
\emph{two} state vectors, together with a new type of physical
quantity, the ``weak value" of a quantum mechanical observable.
Specifically, for a system drawn from an ensemble preselected in
the state $\ket{\psi_1}$ and postselected in the state
$\ket{\psi_2}$, the weak value for the observable $\hat{A}$ is
defined  as
\begin{equation}\label{weakvdef}
A_w \equiv \weakv{\psi_2}{\hat{A}}{\psi_1} \, ,
\end{equation}
where the real part is the quantity of primary physical interest
(and to which the term ``weak value" shall henceforth apply unless
otherwise noted). The suggestion was motivated operationally by
the fact  that both  real and
 imaginary parts of weak
values can be linked to  conditional measurement  statistics
predicted by standard quantum mechanics   for  the general class
of ``weak measurements", defined so as to  minimize the
disturbance to the system  as a result of a diminished interaction
with the measuring instrument.  Under these conditions, joint weak
measurements of two non-commuting observables can be made with
negligible mutual interference, thus ensuring that the
simultaneous assignment of weak values to all elements of the
observable algebra is operationally consistent.

 The usefulness of this  description has been
demonstrated, both theoretically and experimentally, in a number
of applications in which novel aspects of
 quantum processes have been uncovered when analyzed in
terms of weak values. These include  photon polarization
interference\ccite{Duck89,KnightVaid,RSH,Parks99,Brunner03},
barrier tunnelling times \ccite{Stein94,Stein95,AER03}, photon
arrival times \ccite{Ruseckas,Ahnert}, anomalous pulse
propagation\ccite{RA02,Solli04,Brunner04}, correlations in cavity
QED experiments\ccite{Wise02}, complementarity in ``which-way"
experiments\ccite{Wise03,Garretsonetal}, non-classical aspects of
light\ccite{JohNC1,JohNC2}, communication protocols\ccite{BR00}
and retrodiction ``paradoxes" of quantum
entanglement\ccite{ABPRT01,Molmer01,RLS03}.

A certain amount of
skepticism\ccite{Leggett,Peres,AVReply,Kastner98,AVKasreply,Kastner03}
has nevertheless  prevailed regarding the physical status of weak
values, particularly in the light of  the unconventional range of
values that is possible according to\rref{weakvdef}. Indeed, the
real part of $A_w$, describing the ``pointer variable" response in
a weak measurement, may lie outside the bounds of the spectrum of
$\hat{A}$. Manifestly ``eccentric" weak values, as are negative
kinetic energies \ccite{APRV93,RAPV95} or negative particle
numbers \ccite{ ABPRT01,RLS03}, are not easily reconciled with the
physical interpretation that is traditionally attached to the
respective observables. Less intuitive yet is when $\hat{A}$
stands for a projection operator, in which case the  weak value
suggests ``weak probabilities" taking generally non-positive
values\ccite{Stein95,Wang,Garretsonetal}. Such bizarre
interpretations call for a sharper clarification of what physical
meaning  should be attached to the weak value of an observable.

Another item of skepticism surrounding the physical significance
of weak values has to do with their general domain of
applicability. It seems reasonable to demand of any new physical
concept that it be
 applicable to a wide variety of situations outside the
 restricted context in which it is defined operationally.
Although progress has been made in this
direction\ccite{Vaidman96b,Adiabatic}, convincing evidence of the
general validity of the concept of the weak value is still
lacking.

With these questions in mind, the  aim of this paper is two-fold:
First, we address the physical meaning of weak values by showing
that there  exists an unambiguous  interpretation of the real part
of the weak value as a \emph{definite} mechanical effect of the
system on a  measuring probe that is specifically designed to
minimize the dispersion in the back-reaction on the system.
Second, based on this interpretation, we present a new framework
for the analysis of general von Neumann measurements, in which the
measurement statistics are interpreted as {\em quantum averages of
weak values} (QAWV). We  believe this framework is physically
intuitive and provides compelling evidence for the ubiquity of
weak values in more general measurement contexts. In particular,
we show that for arbitrary system ensembles, the expectation value
of the reading of any von Neumann-type measurement is an average
of weak values over a suitable posterior probability distribution.
We  furthermore show how QAWV framework has a natural
correspondence in the classical limit with the posterior analysis
of measurement data according to the classical inference
framework. Thus, we can establish a correspondence between weak
values and what in the macroscopic domain are regarded as
objective classical dynamical variables.

The paper is structured as follows: In Sec. \ref{eigvvswv}, we
motivate the idea of averaging weak values by
 discussing the connection between  pre-selected and pre- and postselected
statistics in arbitrary  measurements von-Neumann type
measurements.  In Sec. \ref{mechint} we present the operational
definition of the weak value as a definite mechanical effect
associated with
 infinitesimally uncertain unitary transformations.
 The QAWV framework  is then
introduced in Sec. \ref{qaves} for arbitrary strength
measurements. We provide an illustration in   Section
\ref{likecc}, where we discuss a number of   measurement
situations in which the framework gives a simple characterization
of the outcome statistics.
%Next, we show in
%Sec. \ref{collective} how in a large-sample setting, it should be
%possible to ``visualize" in terms of frequency distributions, the
%relation between quantum averages of weak values and the posterior
%probability of the back-reaction parameter; in particular, we find
%that that in the large $N$ limit, the quantum average of weak
%values coincides with a sample average of definite weak values.
Finally, we establish in Sec. \ref{Clas} the classical
correspondence of the QAWV framework. Some conclusions are given
in Sec. \ref{concl}.

\section{ Pre- and  Postselected Measurement Statistics, Eigenvalues and Weak Values}
\label{eigvvswv}

The conventional interpretation of a quantum mechanical
expectation value, such as $\qave{\psi}{\hat{A}}$ for an
observable $\hat{A}$, is as an average of the eigenvalues of
$\hat{A}$ over a probability distribution that is realized in the
context of a complete strong measurement of $\hat{A}$. Our main
suggestion in this paper is  that for a wide class of generalized
conditions on the von Neumann measurement
   of $\hat{A}$, the
statistics of measurement outcomes  can alternatively be related
to an underlying statistics of a different quantity,  the weak
value of $\hat{A}$, which is to be regarded as a definite physical
property of an unperturbed quantum system in the time interval
between two complete measurements. We shall therefore begin by
discussing in this preliminary section the connection between pre-
and pre-and post-selected measurement statistics of arbitrary
strength von Neumann measurements, and from this  discussion show
an instance in which averages of weak values more aptly describe
the posterior break-up of the measurement outcome distribution.

In the  von Neumann measurement scheme\ccite{VonNeum}, the device
is some  external system, described by canonical variables
$\hat{q}$ and $\hat{p}$, with $[\hat{q},\hat{p}]=i$ ($\hbar\equiv
1$). The system-device interaction is designed so that the
measurement result is read-off from the effect on some designated
device ``pointer variable", which we take to be $\hat{p}$.  For a
measurement of the system observable $\hat{A}$ at the time
$t=t_i$,  this interaction is modelled by the impulsive
Hamiltonian
\begin{equation}\label{measham}
\hat{H}_m = -  \delta(t-t_i)  \hat{A} \hat{q}  \, .
\end{equation}
(Note that a possible coupling constant can always be absorbed by
canonically redefinig $q$ and $p$.) The effect of the measurement
 is then described by the unitary operator
 $
 \hat{U} = e^{ i   \hat{A} \hat{q} }\,
$.
 Since we will only be concerned with the
effect of this interaction from times immediately before to
immediately after $t_i$, we shall henceforth assume the all
additional free evolution is already contained in the states.

We first consider the pointer variable statistics from an ensemble
defined by pure initial conditions on the system and the
apparatus, described by states $\ket{\psi_1}$ and $\ket{\phi}$,
respectively. For later convenience, we shall term this ensemble
 the \emph{preselected measurement ensemble} (PME) $\Omega_1$.
Further, we introduce the  notation $\prec$ or $\succ$ to denote
times immediately before or immediately after the measurement time
$t_i$. Now, for the PME  $\Omega_1$, the effect of the measurement
interaction is easily described by the  Heisenberg picture
transformation
\begin{equation}
\hat{p}_\aft = \hat{p}_\bef + \hat{A} \, .
\end{equation}
induced by the evolution operator  $e^{ i \hat{A} \hat{q} }$.
Since the initial system plus apparatus state  is separable, the
final statistics of the pointer variable are easily obtained from
the spectral decomposition of $\hat{A}$, and are given by the
probability distribution
\begin{equation}\label{prestats}
{\cal P}(p|\Omega_{1}^\aft ) = \sum_a \langle \pin
|\hat{\Pi}_a|\pin \rangle {\cal P}(p-a|\phi)\, ,
\end{equation}
where $ {\cal P}(p|\phi) =|\langle p |\phi\rangle|^2$,  and
$\hat{\Pi}_a$ is the projector onto the eigenspace of the system
Hilbert space with eigenvalue $a$. In this description, a
``strong" or projective measurement corresponds to the limit
$\Delta p \rightarrow 0$, (i.e., ${\cal P}(p|\phi) \rightarrow
\delta(p)$), in which case  the pointer distribution mimics the
spectral distribution of the Born interpretation, $\langle \pin
|\hat{\Pi}_a|\pin \rangle$. Note however that even if the spectrum
cannot be resolved,  the resulting expression\rref{prestats} for
the pointer statistics can still be interpreted as if, on every
single trial, the pointer variable is displaced in proportion to
one of the eigenvalues of $\hat{A}$, with the eigenvalues
distributed randomly throughout the sample according to  $\langle
\pin |\hat{\Pi}_a|\pin \rangle$. Thus, \emph{regardless} of the
form of ${\cal P}(p |\phi)$ the mean and variance of the
distribution \rref{prestats} will always satisfy
\begin{eqnarray}
\ave{p}_{_{\Omega_{1}^\aft}} & = &   \qave{\pin}{\hat{A}}\, \\
\ave{\Delta p^2}_{_{\Omega_{1}^\aft}} & = & \ave{\Delta p^2}_\phi
+
 \qave{\pin}{\Delta \hat{A}^2}\, \, ,
\end{eqnarray}
where $\ave{\Delta p^2}_\phi $ is the variance in $p$ of the state
$\ket{\phi}$ and we have assumed $\ave{p}_\phi \equiv
\qave{\phi}{p} = 0$ for simplicity.

\begin{figure}
   \epsfxsize=2.65truein
\centerline{\epsffile{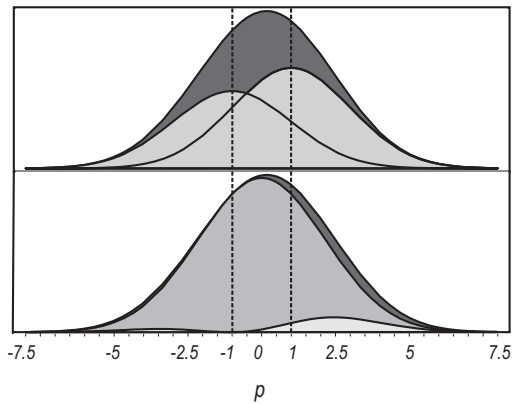}} \caption[\, ]{Decomposition
of event distribution (dark shading) for a $\hat{\sigma}_z$
measurement from a PME defined by a gaussian apparatus state with
$\Delta q = \pi$ and an initial eigenstate of $
\hat{\vect{\sigma}}\cdot \vect{n}_1$ with eigenvalue $+1$,
$\vect{n}_1 = (\sin\theta,0,\cos\theta)$, and   $\theta = 11 \pi/
24$. Top: decomposition according to Eq.\rref{prestats}; Bottom:
actual decomposition in terms of event distributions from the two
PPME's defined by a post selection measurement of
$\hat{\sigma}\cdot \vect{n}_2$, where $\vect{n}_2 =
(-\sin\theta,0,\cos\theta)$. The event distribution for the
unlikely final outcome $\hat{\sigma}\cdot \vect{n}_2=+1$ (lightest
shading) is overwhelmingly concentrated beyond the range of the
spectrum of $\sigma_z$, indicated by dotted vertical lines.
}\label{fig:poolings}
 \end{figure}

Now suppose that after  time $t_i$, a postselection
 is performed on the system, and we wish to concentrate
on the subset of measurement outcomes arising only from those
systems that ended up in some specific state, $\ket{\psf}$. This
final condition defines for us a subensemble $\Omega_{12}$ of the
PME $\Omega_1$, that we call a \emph{pre- and post- selected
measurement ensmble} (PPME), the measurement  statistics of which
can be obtained from the conditional final state of the
apparatus\cite{AV90}
\begin{equation}\label{statetransf}
\ket{\tilde{\phi}_{_{12}}^\aft} = \frac{1}{\sqrt{\psub{12}(\phi)}}
\langle{\psf}|e^{ i \hat{A} \hat{q}}|{\pin}\rangle \ket{\phi}\, ,
\end{equation}
where the normalization $\psub{12}(\phi)$ is shorthand for the
transition probability ${\cal P}(\psf|\pin\phi)$ (i.e.,  the
average relative size  of the ensemble $\Omega_{12}$). From this
state, the corresponding pointer variable distribution  is given
by ${\cal P}(p|\Omega_{12}^\aft) = | \tilde{\phi}_{_{12}}^\aft
(p)|^2$.

Let us briefly discuss some relations between the PME and PPME
statistics. Suppose the postselection involves a complete
measurement of some non-degenerate observable $\hat{B}$, with
eigenstates $ \{ \ket{b} \} $.  A pooling of  the data from all
the subensembles $\{ \Omega_{1b} \}$ of  $\Omega_1$  must then
yield the preselected distribution (Eq.\rref{prestats}), in other
words
\begin{equation}\label{pooldist}
{\cal P}(p|{\Omega_{1}^\aft}) = \sum_b \psub{1 b}(\phi){\cal
P}(p|{\Omega_{1b}^\aft}) \, ,
\end{equation}
where $\psub{1 b}(\phi)$ is the relative size of each PPME. Two
important consequences follow from this decomposition: First, the
PME expectation value of the pointer
$\ave{p}_{_{\Omega_{1}^\aft}}$ breaks up in a similar fashion as
$\ave{p}_{_{\Omega_{1}^\aft}} = \sum_b \psub{1
b}(\phi)\ave{p}_{_{\Omega_{1b}^\aft}}$; assuming that the prior
expectation value of $p$ vanishes, this entails the sum rule
\begin{equation}\label{avesumrule}
\qave{\pin}{\hat{A}} = \sum_b \psub{1
b}(\phi)\ave{p}_{_{\Omega_{1b}^\aft}}\, ,
\end{equation}
i.e, the weighted average of the PPME pointer expectation values
has to yield the standard expectation value of $\hat{A}$. A second
consequence of \rref{pooldist} is a ``covering" condition
satisfied by the individual PPME distributions,
\begin{equation}\label{covercond} {\cal
P}(p|{\Omega_{1}^\aft}) \geq \psub{1 b}(\phi){\cal P}(p|{\Omega_{1
b}^\aft}) \, ;
\end{equation}
for all values of $p$ and all final outcomes$b$. This imposes a
constraint on how rare a PPME $\Omega_{1b}$ should be  were the
corresponding ${\cal P}(p|{\Omega_{1 b}^\aft})$ to be peaked
somewhere in the tail region of ${\cal P}(p|{\Omega_{1}^\aft})$.

The relevance of \rref{avesumrule} and \rref{covercond} is that
  indeed the weight of a  PPME measurement outcome
 distribution  need not lie within the
 ``normal" region of expectation defined by the bounds of the
 spectrum of $\hat{A}$  (Fig.\ref{fig:poolings}), contrary to what one would
 have naively expected given the
 generality of the spectral expansion of the PME distribution\rref{prestats}. This may not be
 obvious under strong measurement conditions,  such that  when
 the appartus wave function in  $p$ is
expanded as a superposition of shifted wave functions
\begin{equation}\label{specexp}
\tilde{\phi}_{_{12}}^\aft(p) \propto \sum_a
\langle{\psf}|\hat{\Pi}_a|{\pin}\rangle   \phi(p-  a)\,  ,
\end{equation}
the overlap between two shifted functions $\phi(p-  a)$ and
$\phi(p- a')$ for all $a \neq a'$ is negligible. Indeed, in such a
case,
 the resulting p.d.f. for the pointer variable  takes the form
of a mixture of ``strong" measurement distributions, i.e.,  $
{\cal P}(p|{\Omega_{12}^\aft}) \propto
\sum_b|\langle{\psf}|\hat{\Pi}_a|{\pin}\rangle|^2\, {\cal P}(p-
a|\phi) $,  each centered at one of the eigenvalues of $\hat{A}$
with weights given by the  Aharonov, Bergmann and Liebowitz rule
for projective measurement sequences\ccite{ABL}; the weight of
this distribution is, of course, within the bounds of the spectrum
of $\hat{A}$. However, away from strong measurement conditions
${\cal P}(p|\Omega_{12}^\aft )$ will involve interference terms
between the shifted wave functions $\phi(p- a)$ with coefficients
$\langle{\psf}|\hat{\Pi}_a|{\pin}\rangle$ that are not generally
real nor positive-definite, preventing the resolution of the
individual shifted peaks and  allowing for destructive
interference effects that may place the weight of
$\tilde{\phi}_{_{12}}^\aft(p)$ beyond the spectrum of $\hat{A}$.

For a wide class of wave functions of the apparatus, weak values
emerge from the limiting behavior of these interference effects in
a complementary limit to that of strong measurement
conditions\ccite{AV90}, namely when $q$, the conjugate to the
pointer variable, satisfies $\Delta q \rightarrow 0$ ($\Delta p
\rightarrow \infty$).   In particular, if $\ave{q}=0$, one obtains
the weak value as the limiting conditional expectation value
\begin{equation}\label{genweakvalapproach}
\lim_{\Delta q \rightarrow 0} \langle p
\rangle_{_{\Omega_{12}^\aft}} \rightarrow {\rm
Re}\weakv{\psf}{\hat{A}}{\pin}\, .
\end{equation}
This limiting behavior is furthermore accompanied by the limit
$\psub{12}(\phi)\rightarrow |\amp{\psf}{\pin}|^2 \, $, as if
indeed no measurement had taken place, justifying the term ``weak
limit". Hence,  in this limit the posterior
break-up\rref{pooldist} of
 the PME pointer distribution is essentially that of a mixture of
 distributions, each of which is centered at the weak value
defined by its corresponding final state in the post-selection and
weighted by the corresponding (unperturbed) transition
probability. Thus we have an instance in which the expectation
value of $\hat{A}$ is more appropriately interpreted operationally
as an average of weak values than as an average of eigenvalues.
Indeed, it is easily verified that the weighted average of the
weak values defined by a complete post-selection is  the standard
expectation value of $\hat{A}$
\begin{equation}\label{avewvs}
 \sum_b  |\amp{\psv{b}}{\pin}|^2\ {\rm Re}
\weakv{\psv{b}}{\hat{A}}{\pin} = \qave{\pin}{\hat{A}}\, ,
\end{equation}
as expected from the sum rule \rref{avesumrule}. Note that this is
a \emph{classical} averaging process, as it arises from the mixing
of the  distributions  conditioned on the distinguishable outcomes
of the post-selection.

The sum rule \rref{avewvs} embodies a general rule of thumb,
namely that \emph{``eccentric weak values are unlikely"},
according to which weak values lying outside the spectrum of
$\hat{A}$ must be weighted by correspondingly small relative
probabilities, ensuring that the average over all pre- and
postselected  subensembles yields a quantity within the spectral
bounds of $\hat{A}$.  This generic property of weak values is at
the heart of the QAWV framework presented in Section \rref{qaves},
where we show that pre-and post-selected statistics away from weak
measurement conditions can also be interpreted from a
\emph{quantum} averaging process involving  weak values.

\section{Mechanical Interpretation of Weak Values}
\label{mechint}

Implicit in the suggestion that standard expectation values can be
interpreted (at least under certain conditions) as averages of
weak values,  is the idea that weak values are in some sense
``sharp" physical properties. We therefore expand on this notion
of ``sharpness" by giving an operational sense in which the weak
value can indeed be regarded as a definite mechanical property of
a system that is known to belong to an enesemble defined by
complete pre- and post-selections.

The functional dependence on  $q$ of the transition amplitude
$\langle{\psf}|e^{ i q \hat{A} }|{\pin}\rangle$ in
\rref{statetransf}  furnishes the necessary elements to build a
 description of the PPME statistics based on a picture
 of ``action and reaction", in which,
  if the variable $q$ is sharply defined, then a)  the
measured system is subject to a sharply-defined unitary
transformation generated by $\hat{A}$, and b) the measuring
apparatus suffers a sharply-defined response given by the weak
value of $\hat{A}$. This elementary picture serves the basis for
the more general QAWV framework discussed in the following
section.

Let us look at the polar decomposition of $\langle{\psf}|e^{ i q
\hat{A} }|{\pin}\rangle$, which we choose to express as
\begin{equation}
\label{polardecomp}  \langle{\psf}|e^{ i   \hat{A}
q}|{\pin}\rangle =\sqrt{\psub{12}(q)} \, e^{ i S_{\!_{12}}(q) }\,
,
\end{equation}
where
\begin{equation}
\psub{12}(q) \equiv \left|\,\langle{\psf}|e^{ i \hat{A}
q}|{\pin}\rangle\right|^2 \,
\end{equation}
(see also refs. \ccite{Botero03,Botero04,Solli04}) gives the
transition probability from $\ket{\pin}$ to  $\ket{\psf}$, but
mediated by an intermediate unitary transformation $e^{ i \hat{A}
q}$.
%From the
%modulus, we see that $\psub{12}(q)$ is the transition probability
%between the states $\ket{\pin}$ and $\ket{\psf}$, but mediated by
%an intermediate unitary transformation  $e^{i \hat{A} q}$.
Thus, the variable $q$ can be regarded as the parameter of a
back-reaction on the system, generated by the operator $\hat{A}$,
inducing the transformation of the initial state
\begin{equation}\label{backreact}
\ket{\pin} \stackrel{q}{\rightarrow} \ket{\pin(q)} \equiv e^{i
\hat{A}
 q}\ket{\pin}\,
\end{equation}
(alternatively, the reaction can be viewed as the inverse
transformation $e^{-i \hat{A} q}$ on the final state
$\ket{\psf}$).
%Thus,  variations in $q$ entail a sampling on the
%orbit of transformed initial states $\ket{\pin(q)}$. This sampling
%is reflected, for instance, in the transition probability
%$\psub{12}(\phi)$ obtained from the normalization of
%$\ket{\tilde{\phi}_{_{12}}^\aft}$; since this probability can be
%expressed as the integral over $q$,
%\begin{equation}\label{psub}
%\psub{12}(\phi) = \int dq\, \psub{12}(q) {\cal P}(q|\phi) \, ,
%\end{equation}
%it may therefore be interpreted as the  average of the transition
%probabilities between all possible rotated states $\ket{\pin(q)}$
%and the final state $\ket{\psf}$, weighted by the prior
%probability of the  reaction parameter  $q$, ${\cal P}(q|\phi)$.
On the other hand, the phase factor in\rref{polardecomp} can be
viewed as the generator of a certain reaction  of the system on
the apparatus corresponding to a specific rotation parameterized
by $q$:  viewed as a unitary operator on the apparatus degrees of
freedom, $e^{ i S_{\!_{12}}(\hat{q})}$ induces in the Heisenberg
picture  the generally nonlinear canonical transformation of the
pointer operator
\begin{equation}\label{canshift}
\hat{p}_\aft = \left. e^{- i S_{\!_{12}}(\hat{q}) } \hat{p}\, e^{i
S_{\!_{12}}(\hat{q}) }\right|_\bef \equiv \hat{p}_\bef +
\wva{12}(\hat{q})\, ,
\end{equation}
where
$
\wva{12}(q) \equiv S'_{_{12}}(q) ={\rm Im} \frac{d }{dq}
\log\langle{\psf}|e^{ i   \hat{A} q}|{\pin}\rangle \, .
$
A straightforward derivation then shows that $\wva{12}(q)$ is
indeed a weak value
\begin{equation}
\wva{12}(q) = {\rm Re} \frac{\bra{ \psf}{\hat{A}e^{i \hat{A}
q}}\ket{\pin}}{\bra{ \psf}{e^{i \hat{A} q}}\ket{\pin}}={\rm
Re}\weakv{\psf}{\hat{A}}{\pin(q)}\, ,
\end{equation}
namely the weak value of $\hat{A}$ for the rotated state
$\ket{\psi_i(q)}$ and the final state $\ket{\psf}$.
Equation\rref{canshift} therefore shows that for a definite value
of $q$, there is an associated definite reaction on the measuring
device pointer variable by the weak value for the corresponding
pair of states $(\ \ket{\psi_i(q)}\, ,\, \ket{\psf}\ )$.

More precisely, note that for the general pointer variable
statistics for the PPME $\Omega_{12}$, Eq.\rref{specexp}, we can
equivalently express the final apparatus wave function
$\tilde{\phi}_{_{12}}^\aft(p)$ as the  Fourier integral
\begin{equation}\label{fourint}
\tilde{\phi}_{_{12}}^\aft(p) = \frac{1}{\sqrt{ 2
\pi}}\int_{-\infty}^{\infty} d q\,
\sqrt{\frac{\psub{12}(q)}{\psub{12}(\phi)}}\phi(q)e^{-i [p
q-S_{_{12}}(q) ] }\, .
\end{equation}
Let us now suppose that  $q$ is constrained to lie exclusively
within a finite range around some value $q = q_i$, by taking
$\phi(q)$ to be the ``window" function of width $\varepsilon$
centered at $q = q_i$.
\begin{equation}\label{weaktrans}
W_{q_i, \varepsilon}(q) =  \left \{ \begin{array}{ccc}
\frac{1}{\sqrt{ \varepsilon}} \, , & &
 |q - q_i| < \frac{\varepsilon}{2} \\ 0\, , &  & |q - q_i| \geq
 \frac{\varepsilon}{2}\end{array}\right. \, .
\end{equation}
In this case  the wave function in the $p$-representation is a
modulated ``sinc" function
\begin{equation}
 W_{q_i, \varepsilon}(p) = \sqrt{\frac{2}{\varepsilon \pi}}\, \frac{\sin\left( \frac{\varepsilon
p}{2}\right)}{p}e^{i p q_i} \, ,
\end{equation}
of characteristic width $\sim 1/\varepsilon$. Now let
$\varepsilon$ be small enough that  variations of $\psub{12}(q)$
and $\wva{12}(q)$ are negligible within the interval $|q - q_i| <
\frac{\varepsilon}{2}$. Thus, we can  approximate $\psub{12}(\phi)
\simeq \psub{12}(q_i)$, and perform the Fourier integral  in the
``group velocity approximation", i.e., by expanding the phase
about $q_i$ to first order and replacing $\psub{12}(q)$ by
$\psub{12}(q_i)$; this yields
\begin{equation}\label{mechdef}
\tilde{\phi}_{_{12}}^\aft(p) \simeq e^{i
\Gamma_{\!_{12}}(q_i)}W_{q_i, \varepsilon}(p-\wva{12}(q_i)) \, ,
\end{equation}
where we define $ \Gamma_{\!_{12}}(q) \equiv S_{\!_{12}}(q) - q\,
\wva{12}(q) \, . $ Hence, in the limit $\varepsilon \rightarrow
0$, where the apparatus wave function approaches an eigenstate of
$\hat{q}$ with eigenvalue $q_i$, the final wave function for the
pointer becomes (up to a phase) the initial wave function
\emph{rigidly shifted by a definite weak value}, the weak value
$\wva{12}(q_i)$ for the rotated state $\ket{\pin(q_i)}$ and the
final state $\ket{\psf}$.

From the point of view of the system, the limit $\varepsilon
\rightarrow 0$  can be regarded as an idealization of  a situation
often encountered in more general contexts, where the evolution of
a quantum system is treated as effectively unitary despite the
fact that  certain parameters of the evolution are actually
physical variables of some external (and typically macroscopic)
system; for example  a spin rotation, where a macroscopic external
magnetic field sets the rotation angle. That such parameters can
be treated as classical numbers is a consequence of a negligible
uncertainty of the  quantum variable of the external system acting
as the parameter for the transformation. The interaction with such
an external system may thus be idealized as an
\emph{infinitesimally uncertain unitary transformation}  at a
given parameter value. This idealization provides the desired
mechanical definition of weak values:
 \emph{The weak value $\wva{12}(q)$ corresponds to a
definite conditional reaction of the system on the variable
conjugate to the external physical ``parameter variable" $\hat{q}$
of  an infinitesimally uncertain unitary transformation generated
by $\hat{A}$ at parameter value $q$}. The essence of a weak
measurement is thus to approach, as close as possible, the ideal
conditions of an infinitesimally uncertain transformation.

The above  definition presents no ambiguities in the physical
interpretation  of ``eccentric" weak values or in the sometimes
unexpected relationships that may arise between the weak values of
say, $\hat{A}$ and $\hat{A}^2$ (e.g., negative ``weak variances",
etc.). To the extent that we associate
 weak values  to  infinitesimal unitary
transformations, no \emph{a-priori} connection between the weak
values of two commuting observables should be expected; typically,
commuting operators such as $\hat{A}$ and $\hat{A}^2$ generate
entirely different types of un unitary transformations. Rather,
relations between weak values follow from the \emph{linear},
vector space structure of the Lie Algebra of  hermitian operators
generating infinitesimal transformations.  The vector space
structure is reflected, for instance, in the fact that for any two
initial and final states that are eigenstates of the the
observables $\hat{A}$ and $\hat{B}$, with eigenvalues $a$ and $b$
respectively, the reaction to an infinitesimal unitary
transformation generated by the linear combination
$\hat{C}\equiv\alpha \hat{A} + \beta \hat{B} $  at $q=0$  is the
linear combination $\mathcal{C}=\alpha a + \beta b$.

Finally, let us emphasize the significance of the present
mechanical interpretation of weak values in connection with
certain quantum mechanical operators, such as kinetic energy or
particle number\ccite{AV91,APRV93,ABPRT01}, for which any
association with negative values would appear to be forbidden. The
fact that the
 reactions associated with weak values will almost
always lie within the
 range of the  observable's spectrum is what gives us a reference from which to
 identify, in
those unlikely circumstances
 where the reaction is ``eccentric",
 what are unique quantum-mechanical effects associated with the role of
 the observable as a generator of infinitesimal transformations.
One would hardly suspect that such effects could indeed be
possible   given the physical interpretations that we have
traditionally attached  to the eigenvalues of a quantum mechanical
observable.

\section{Quantum Averages of Weak Values}
\label{qaves}

The framework of quantum averages of weak values (QAWV) is the
extension of the previous analysis to  general von Neumann
measurements, with arbitrary pure initial states of the apparatus
not necessarily satisfying a ``weakness condition". Given a
\emph{pure} PPME $\Omega_{12}$,  we shall show how the conditional
average of measurement outcomes can nevertheless be interpreted as
a quantum average of weak values over a suitable distribution.
 For more general initial and final conditions on the system (as
 well as more general initial conditions on the apparatus),
 the corresponding averages can then be obtained by a
 classical averaging process, similar to that of \rref{avewvs}, given that any such ensemble  can always be
broken-up into complete pre-and postselected measurement
subensembles with appropriate relative weights.

The heuristics of the framework are straightforward: a general
apparatus pure state $\ket{\phi}$ entails indefiniteness in the
parameter value $q$ driving the back-reaction on the system
according to \rref{backreact}, so that a generally finite range of
system configurations are sampled in the orbit of transformed
initial states $\ket{\pin(q)}$. Correspondingly, the pointer
measurement statistics should reflect the sampling of a certain
range of weak values $\wva{}(q)$ associated to this orbit.
However, once $q$ is allowed to take arbitrary values, a new
element in the description comes into play. This has to do with
the  relative weights associated with the sampled values of $q$,
which reflect a probability-reassessment in the light of the
additional conditions entailed by the post-selection. The central
idea of the framework is then that an arbitrary strength von
Neumann measurement on a pre- and postselected system may be
viewed as a superposition of weak measurements at different
sampling points $q$, with a re-assessment of the weights of each
sample in accordance with Bayes' theorem.

Let us for simpliciity consider an initial apparatus function that
is   real and smooth. This function may then be represented as the
limit of a superposition of infinitesimally-wide window functions
\begin{equation}
\phi(q) =\lim_{\varepsilon\rightarrow 0} \sum_{k=-\infty}^{\infty}
\sqrt{\varepsilon} \phi(q_k)W_{q_k, \varepsilon}(q) \, ,
\end{equation}
centered at the ``sampling points" $q_k = k  \varepsilon + \delta$
with $k \in \mathbb{Z}$ and $\delta \in [-\varepsilon/2,
\varepsilon/2)$. From the results of the previous section, and by
linearity, the corresponding final apparatus state wave function
may be represented as
\begin{equation}
\tilde{\phi}_{_{12}}^\aft(p) =\lim_{\varepsilon\rightarrow 0}
\sum_{k=-\infty}^{\infty}
 \sqrt{\varepsilon\frac{\psub{12}(q_k)}{\psub{12}(\phi) }}
\phi(q_k)e^{\Gamma_{\!_{12}}(q_k)}W_{q_k,
\varepsilon}(p-\wva{12}(q_k)) \, .
\end{equation}
The final apparatus state can therefore be viewed as a
superposition of weak measuerements at the sampling points $q_k$
 but with the initial
weights $\phi(q_k)$ replaced by new weights $\sqrt{
{\psub{12}(q_k)}/{\psub{12}(\phi) }}\phi(q_k)$. As is easily seen,
this re-assessment of weights is in correspondance with ${\cal
P}(q|\Omega_{12})$, the  p.d.f. for strong
 measurements
  of $q$
   (performed either before or after the measurement interaction) on the PPME
$\Omega_{12}$. Consistently with Bayes' theorem, ${\cal
P}(q|\Omega_{12})$ is the posterior distribution for $q$ after  a
re-assessment of the prior p.d.f. ${\cal P}(q|\phi)$ by the
likelihood $\psub{12}(q)/\psub{12}(\phi)$ of the post-selection
given the $q$-dependent rotation of the initial state:
\begin{equation}\label{postdef}
{\cal P}(q|\Omega_{12}) = \frac{\psub{12}(q)}{\psub{12}(\phi) }\,
{\cal P}(q|\phi) \, .
%|\langle{\psf}|{\pin}(q)\, \rangle|^2 |\phi(q)|^2 \, .
\end{equation}
 Note that in accordance with the ``eccentric weak values
are unlikely" rule of thumb,  the likelihood factor $\propto
\psub{12}(q)$ will tend to suppress the contributions in the
superposition  for which the weak value falls outside the spectrum
of $\hat{A}$. As we shall illustrate in the coming section, it is
this  mechanism that ensures, together with quantum mechanical
interference, that the strong measurement distributions peaked at
the eigenvalues of $\hat{A}$ can nevertheless be understood as a
quantum superpositions of weak measurements.

It becomes convenient to  capture in compact form the two
conceptually different processes involved in the updating of the
apparatus state $\ket{\phi} \rightarrow
\ket{\tilde{\phi}_{_{12}}^\aft}$ as a result of the measurement.
The first step, the generally irreversible process of probability
re-assessment, can be expressed conveniently by defining a
fiducial state $\ket{\tilde{\phi}_{_{12}}^\bef}$, which we term
the \emph{re-assessed initial state} of the apparatus. Defining
the state  by its wave function in $q$, it corresponds to the
(prior) initial  wave function $\phi(q)$ multiplied by the square
root of the likelihood factor in\rref{postdef}:
\begin{equation}\label{statereassess}
\tilde{\phi}_{_{12}}^\bef(q) =
\sqrt{\frac{\psub{12}(q)}{\psub{12}(\phi) }}\, \phi(q) \, .
\end{equation}
The other process is the  reversible mechanical action of the
system on the measurement apparatus generated by the unitary
operator $e^{i S_{\!_{12}}(\hat{q})}$ defined by the polar
decomposition \rref{polardecomp}.  The final conditional state of
the measuring device  can then be expressed as a unitary
transformation applied to the re-assessed state
$\ket{\tilde{\phi}_{_{12}}^\bef}$, $
\ket{\tilde{\phi}_{_{12}}^\aft} = e^{i S_{\!_{12}}(\hat{q})
}\ket{\tilde{\phi}_{_{12}}^\bef} $. Equivalently, one can compute
pointer statistics  from the Heisenberg picture
transformation\rref{canshift}  using the apparatus state
$\ket{\tilde{\phi}_{_{12}}^\bef}$. In particular, the conditional
p.d.f. of pointer readings can be
 expressed as a quantum-mechanical analogue of a marginal
distribution of shifted pointer values,
\begin{equation}\label{quantpdf}
{\cal P}(p|\Omega_{12}^\aft) = \left \langle
\tilde{\phi}_{_{12}}^\bef \biggl| \, \delta\!\bigl(p - \hat{p} -
 \wva{12}(\hat{q})\,
\bigr)\biggr|\tilde{\phi}_{_{12}}^\bef\, \right\rangle \, ,
\end{equation}
in other words, as a \emph{quantum average of weak values}, where
the average is taken with respect to the re-assessed initial state
$\ket{\tilde{\phi}_{_{12}}^\bef}$.  As we shall see in section
\ref{Clas},  Eq.\rref{quantpdf} has a natural correspondence in
the classical limit;  it can be shown to correspond with the
marginal posterior p.d.f. for the measurement outcomes of the
classical function corresponding to $\hat{A}$ on a classical
canonical system specified by initial and final boundary
conditions in time.

Using the Heisenberg picture, we finally obtain the   pointer
reading mean and variance for the PPME $\Omega_{12}$
\begin{subequations}\label{finalmoms}
%\begin{eqnarray}
%\ave{p}_{_{\Omega_{12}^\aft}} & = &  \ave{\wva{12}}_{_{\Omega_{12}}} \, \label{meaneq}\\
%\ave{\Delta p^2}_{_{\Omega_{12}^\aft}} & = &
%\qave{\tilde{\phi}_{_{12}}^\bef}{ \hat{p}^2} +  \ave{(\Delta
%\wva{12})^2}_{_{\Omega_{12}}} \label{vareq}\, ,
%\end{eqnarray}
\begin{eqnarray}
\!\!\!\!\!\!\!\!\!\!\!\!\!\!\langle p \rangle_\aft &\!\!\! =\!\!\!
& \langle p \rangle_\bef + \langle
\wva{12} \rangle_\bef \, \\
\!\!\!\!\!\!\!\!\!\!\! \langle \Delta p^2 \rangle_\aft &\!\!\!
=\!\!\! & \langle \Delta p^2 \rangle_\bef +  \langle \{\Delta p,
\Delta \wva{12}\} \rangle_\bef +  \langle \Delta \wva{12}^2
\rangle_\bef\, ,
\end{eqnarray}
\end{subequations}
where the subscripts $\bef$ and $\aft$ stand  for expectation
values in the state $\ket{\tilde{\phi}_{_{12}}^\bef}$ and
$\ket{\tilde{\phi}_{_{12}}^\aft}$, and where, the quantum weak
value average $ \ave{\wva{12}}$ and variance $\langle \Delta
\wva{12}^2 \rangle_\bef$ are directly evaluated using the
posterior p.d.f. \rref{postdef}. These expressions can be further
simplified if the initial apparatus state has a real
$\phi(q)$\ccite{BoteroThesis,JohArb} and vanishing expectation
value of $p$, in which case the posterior expectation $\langle p
\rangle_\bef$ and the correlation $\langle \{\Delta p, \Delta
\mathcal{A}\} \rangle_\bef$  vanish. Under such conditions,  the
first two central moments of\rref{quantpdf} are indistinguishable
from those obtained from classically averaging weak values with a
variability defined through  the posterior
distribution\rref{postdef}. Note therefore that a condition  for a
weak measurement that is more general than the one discussed in
the previous section is that we have a sharp \emph{posterior}
p.d.f in $q$ around some value $q=q_*$, in which case  the pointer
average reflects  a measurement of a sharply-defined weak value
$\simeq \wva{12}(q_*)$ with a small  uncertainty $\langle \Delta
\wva{12}^2 \rangle_\bef$. Examples of how such effective weak
measurements are attained will be given in the next section.

Equation\rref{quantpdf} provides a statistical characterization of
the pointer variable response as a quantum average of weak values,
given the most restrictive conditions possible for a pre- and
postselected measurement ensemble. Statistics from less
restrictive measurement ensembles can then be obtained using
standard probability assessments on the PPME $\Omega_{12}$
consistent with the specified conditions. In particular,  for the
preselected measurement ensemble $\Omega_1$ and some specific
post-selection measurement, the pointer variable distribution
\rref{quantpdf} obeys equation Eq.\rref{pooldist}. This
correspondence yields a generalization of the sum
rule\rref{avewvs} to arbitrary measurement strengths, involving
both classical and quantum averages
\begin{equation}
 \qave{\pin}{\hat{A}}  =  \sum_b \psub{1 b}(\phi)
\ave{\mathcal{A}_{_{1b}}}_{_{\Omega_{1b}}}\, ,
\end{equation}
and which is easily verified using Eqs.\rref{postdef}
and\rref{avewvs}.   Even more generally, since the statistics for
any set of less restrictive conditions on the system and/or the
apparatus will involve a classical averaging over the states
$\ket{\phi}$, $\ket{\pin}$, and $\ket{\psf}$,  the final
expectation value of any von-Neumann type measurement can always
be connected to a suitable average of weak values.

\section{Illustration of the QAVW Framework}
\label{likecc}

\begin{figure}
   \epsfxsize=2.4truein
\centerline{\epsffile{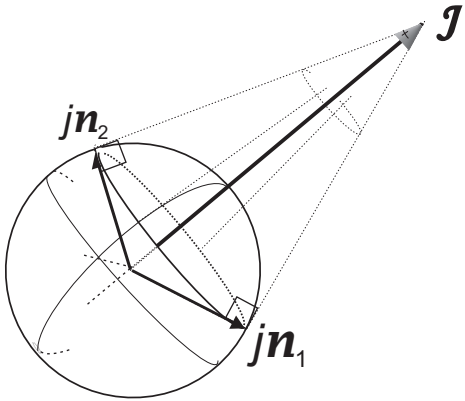}}
\medskip
\caption[\, ]{Geometric construction for the weak spin vector
$\vect{\mathcal{J}}$ for initial and final spin-$j$ coherent
states $\ket{\vect{n}_1;j}$ and $\ket{\vect{n}_2;j}$.
}\label{fig:spin}
 \end{figure}
\begin{figure}
   \epsfxsize=2.0truein
\centerline{\epsffile{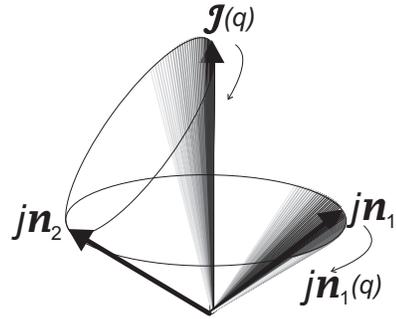}} \caption[\, ]{Orbit of the
weak spin vector for a $\hat{J}_z$ measurement, assuming the $z$
axis coincides with the direction of $\vect{\mathcal{J}}(0)$.
}\label{fig:sl}
 \end{figure}

Let us then illustrate how the  QAVW framework provides new
insight  into the measurement statistics of arbitrary-strength von
Neumann measurements in pre-and post-selected ensembles. A
 particularly graphic example of
the interrelationship between the orbit of weak values
$\mathcal{A}_{\!_{12}}(q)$ and the corresponding likelihood
function $\propto P_{12}(q)$  is that of spin-component
measurements given initial and final spin-$j$ coherent
states\ccite{Perelomov}. Let $\vect{n}$ be a unit vector with
direction parameterized by the polar angles $\theta$ and $\phi$,
and $\ket{j,j}$ the maximal weight $J_z$ eigenstate
($\hat{J}_z\ket{j,j}= j\ket{j,j}$); a spin-$j$ coherent state is
then defined as
\begin{equation}
\ket{\vect{n};j} = e^{ -i \hat{J}_z \phi}e^{ -i \hat{J}_y
\theta}\ket{j,j}\, ,
\end{equation}
 and is hence an eigenstate of $\hat{J}_\vect{n} \equiv
 \hat{\vect{J}}\cdot\vect{n}$. Calculations are  simplified by the fact that this state can be realized as a
  product state of $2j$ copies of the spin-$1/2$ coherent state
 $\ket{\vect{n};\frac{1}{2}}$. In particular, the transition probability between two coherent
 states is
\begin{equation}
|\amp{\vect{n}_2;j}{\vect{n}_1;j}|^2 \propto  ( 1 +
\vect{n}_2\cdot\vect{n}_1  )^{2 j} \, ,
\end{equation}
while the weak value of all spin components are easily captured by
a weak spin vector
\begin{equation}
\vect{\mathcal{J}} \equiv{\rm Re} \weakv{\vect{n}_2;j}{\hat{\vect{J}}}{\vect{n}_1;j}\\
          = j \frac{ \vect{n}_2 + \vect{n}_1}{1
          +\vect{n}_2\cdot\vect{n}_1} \, ,
\end{equation}
for which the projection onto both $\vect{n}_2$ and  $\vect{n}_1$
is $j$ (Fig.~\ref{fig:spin}). Note the relation between
$|\amp{\vect{n}_2;j}{\vect{n}_1;j}|^2$ and the length
$\mathcal{J}$ of the weak spin vector,
$
|\amp{\vect{n}_2;j}{\vect{n}_1;j}|^2 \propto {\mathcal{J}}^{-4 j}
\,
$
in consistency with the  ``eccentric weak values are unlikely"
rule.

\begin{figure}
   \epsfxsize=3.5truein
\centerline{\epsffile{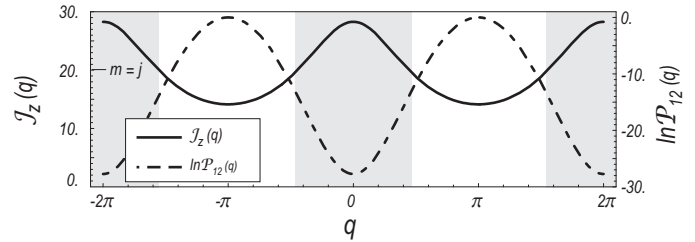}} \caption[\, ]{Weak value
$\mathcal{J}_{z}(q)$ and log-likelihood $\ln \mathcal{P}_{12}(q)$
for $j = 20$ and $\hat{J}_z$ measurement with $\vect{n}_1 =
(0,\frac{1}{\sqrt{2}},\frac{1}{\sqrt{2}})$ and
$\vect{n}_2=(0,-\frac{1}{\sqrt{2}},\frac{1}{\sqrt{2}})$, as
depicted in Fig.~\ref{fig:sl}. Shaded areas represent regions
where the weak value lies outside the bounds of the spectrum
($|\mathcal{J}_{z}(q)|>j$). }\label{fig:spincurves}
 \end{figure}

In a measurement of the spin component $\hat{J}_z$ on a PPME
defined by initial and final coherent states $\ket{\vect{n}_1;j}$
and $\ket{\vect{n}_2;j}$, the back-reaction corresponds to a spin
rotation of the initial state about the $z$-axis by the angle
$-q$:
\begin{equation}
\ket{\vect{n}_1;j} \stackrel{q}{\rightarrow}
\ket{\vect{n}_1(q);j}\, , \ \ \ \vect{n}_1(q) =
R_z(-q)\vect{n}_1\, .
\end{equation}
This reaction in turn entails an orbit for the weak spin vector
$\vect{\mathcal{J}}(q)$ (see Fig~\ref{fig:sl}), from which the the
weak value function $\mathcal{J}_{z}(q)$ for $\hat{J}_z$ can be
obtained by projecting onto the $z$-axis. Note that since the $z$
component of $\vect{n}_1$ is unaffected by the rotation,
$\mathcal{J}_{z}(q) \propto (1 +
\vect{n}_2\cdot\vect{n}_1(q))^{-1}$; thus, the likelihood factor
satisfies
\begin{equation}\label{spinzbias}
\mathcal{P}_{12}(q)   \propto \mathcal{J}_{z}(q)^{-2 j} \, .
\end{equation}
Figure~\ref{fig:spincurves} illustrates the correlated behavior of
$\mathcal{J}_{z}(q)$ and $\ln \mathcal{P}_{12}(q)$ for the case
$j=20$ and with initial and final spin coherent states with
$\vect{n}_1 = (0,\frac{1}{\sqrt{2}},\frac{1}{\sqrt{2}})$ and
$\vect{n}_2=(0,-\frac{1}{\sqrt{2}},\frac{1}{\sqrt{2}})$. For these
conditions,  the weak value is given by
\begin{equation}\label{wvalj}
\mathcal{J}_{z}(q) = j \frac{\sqrt{2}}{1 +
\sin^2\!\left(\frac{q}{2}\right) } \, ,
\end{equation}
oscillating between  $j \sqrt{2}$ at $q_+ = 2n\pi $ (full
rotations of the initial state), and   $j/ \sqrt{2}$ at $q_- =
(2n+1)\pi$ when $\vect{n}_1(q)$ coincides with $ \vect{n}_2$). As
the figure shows, for $j \gg 1$, the likelihood
$\mathcal{P}_{12}(q)$ shows essentially an exponential  behavior
similar to a modular gaussian distribution; in particular, near
values  $q_+$ or $q_-$ (both periodic), for which the magnitude of
$\mathcal{J}_{z}(q)$ is respectively either maximal or minimal on
the orbit, we have the approximations for large $j$
\begin{equation}\label{likeapp}
\mathcal{P}_{12}(q) \simeq \left|\mathcal{J}_{z}(q_\pm)\right|^{-2
j}e^{\pm j
\left|\frac{\mathcal{J}_z''(q_\pm)}{\mathcal{J}_z(q_\pm)}\right|
(q - q_\pm)^2 } \, .
\end{equation}
The exponential suppression of\rref{likeapp} near $q_+$, where the
weak value is maximal in magnitude, is   generic of the
 phenomenon of Fourier
superoscillations\ccite{AAPV90,Berry92,ABRS98,Kempf00,BCG93},
exhibited by  the amplitude $\langle{\vect{n}_2;j}|e^{ i \hat{J}_z
q}|\vect{n}_1;j\rangle$ near $q_{+}$. This suppression imposes a
``robustness" condition on the prior distribution in $q$ if one is
to measure eccentric weak values near $q_+$: not only must the
prior distribution be ``sharp" around $q=q_+$, but additionally it
must show a sufficiently fast fall-off to overcome the exponential
rise in likelihood.

\begin{figure}
   \epsfxsize=3.50truein
\centerline{\epsffile{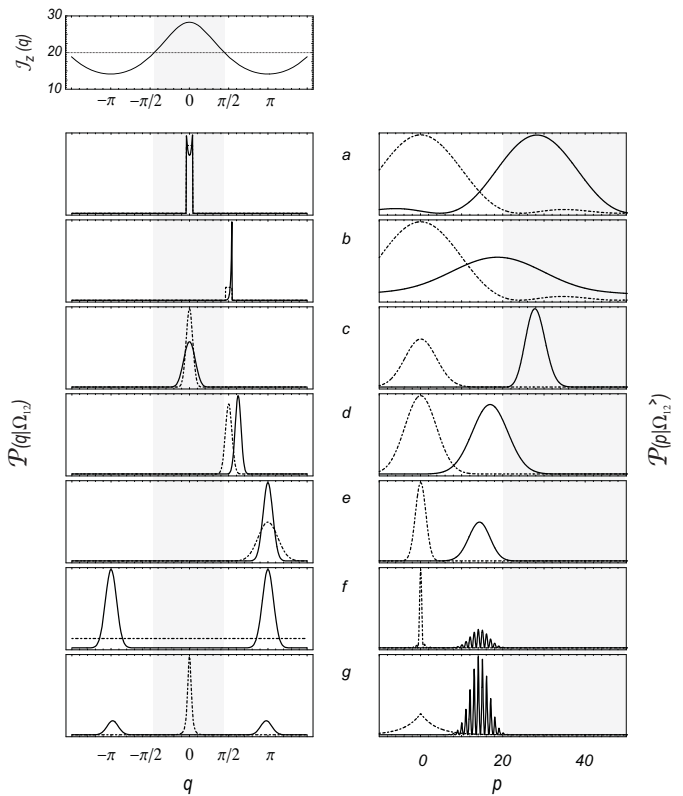}} \caption[\, ]{Interplay
between prior distribution in $q$ and the likelihood factor, and
resulting consequences on the pointer distribution (see text).
Figures a) through g) on the left show prior (dotted) and
posterior (solid) distributions in $q$, while figures on the right
show the corresponding prior (dotted) and final posterior (solid)
distributions in $p$. The top figure shows the curve of the weak
value. Shaded regions correspond to regions of ``eccentric"
effects. }\label{fig:likeffex}
\end{figure}

%\section{Likelihood Effects, Weak to Strong Measurement Transitions}
%\label{likexamp}

On the basis of the generic correlated behaviors of the likelihood
function and the weak value,   the PPME pointer statistics for a
relatively wide range of von Neumann measurement
conditions--ranging from weak to strong measurements--can  easily
be described in the QAWV framework  using  simple sampling
profiles. As discussed in the previous section, a sharp
\emph{posterior} distribution ${\cal P}(q|\Omega_{12})$ about some
well-defined ``sampling point" $q_\post$ satisfies the conditions
for a weak measurement.
%Correspondingly, if $\tilde{\phi}_{_{12}}^\bef(p)$ is
%the re-assessed initial wave function,  the PPME pointer
%distribution for a measurement of $\hat{A}$ with weak measurement
%conditions should be approximately
%\begin{equation}
%{\cal P}(p|\Omega_{12}^\aft) \simeq
%\left|\tilde{\phi}_{_{12}}^\bef(p - \wva{12}(q_\post)) \,
%\right|^2\, .
%\end{equation}
More generally, however, the reassessment by the likelihood factor
of the initial apparatus state $\ket{\phi}$  may yield a state
$\ket{\phi_{_{12}}^\bef}$ for which the wave function in $q$ shows
several  well-separated  narrow peaks, each  satisfying weak
measurement conditions. In this case, the PPME pointer
distribution will be the result of a coherent superposition  of
weak measurement pointer wave functions, and will therefore
exhibit  interference fringes. In simple examples, the existence
of just two peaks may be all that is needed to produce the
statistical distributions associated with strong measurement
conditions (i.e., with maxima at the eigenvalues of the measured
observable).

Figures~\ref{fig:likeffex}a to~\ref{fig:likeffex}g show how such
single or multiple weak measurement conditions are attained from
prior distributions in $q$, of various shapes and locations, for
the    spin-$j$ PPME setting of Fig.~\ref{fig:spincurves} (weak
value ranging between $j/\sqrt{2}$ to $\sqrt{2}j$),
%To understand
%the general effect of the likelihood factor in these examples, we
%find it convenient to define the bias function
%\begin{equation}
%\bias{12}(q) \equiv \frac{d}{dq}\ln \mathcal{P}_{12}(q)=  -2{\rm
%Im}\weakv{\psf}{\hat{A}}{\pin(q)}\, ,
%\end{equation}
%which in turn shows the operational significance of the imaginary
%part of the complex weak value
%$\bra{\psf}{\hat{A}}\ket{\pin}/\amp{\psf}{\pin}$.
 with real wave functions for the initial state of the
apparatus. Starting with Figs.~\ref{fig:likeffex}a and
~\ref{fig:likeffex}b, we illustrate the likelihood effects on an
initial robust state of the apparatus given by a narrow window
function in $q$ of the form given by Eq.\rref{weaktrans}, with two
different locations $q_i$. Such profiles guarantee that $q$, and
hence the average weak value, will always lie within a specific
interval; thus, the effect of the likelihood factor will primarily
be a distortion in the shape of the pointer distribution, with
minimum effect on the expectation value of $p$.
%If $q_i$ is at, or very close, to a
%likelihood minimum ($\bias{12}'=0$) as in
%Fig.~\ref{fig:likeffex}a, and additionally the width of the window
%chosen so that  $\varepsilon \ll 1/\sqrt{\bias{12}'}\,$, the
%rectangular profile is only slightly altered by the likelihood
%factor;  the pointer distribution is then essentially the prior
%distribution $\mathcal{P}(p|\phi)$ shifted by the weak value at
%$q_i$. More generally, if $\bias{12}\neq 0$, and  $\varepsilon \ll
%1/\sqrt{\bias{12}'(q_i)}\,$, the posterior distribution in $q$
%takes the shape $\sim \exp\left[\bias{12}(q_i) q\right]$ biased
%towards the region of increasing likelihood. For
%$\bias{12}\epsilon \ll 1$, the rectangular profile is again only
%slightly distorted; if the ``sampling point" $q_\post$ is  defined
%as the posterior expectation value of $q$, the leading correction
%can be obtained from a linear expansion of the likelihood factor,
%yielding in general
%\begin{equation}\label{linapp}
%q_\post \simeq q_i + \ave{\Delta q^2}_\phi \bias{12}(q_i) +
%O(\Delta q^3)\, ,
%\end{equation}
%($\ave{\Delta q^2}_\phi =\varepsilon^2/{12}$ for the window
%function).  On the other hand, if $\bias{12}(q_i) \gtrsim
%\varepsilon$, as in Fig.~\ref{fig:likeffex}b, the effect on the
%window function can be quite dramatic. In such a case, the
%posterior distribution in $q$ takes a sharp needle-like shape with
%$q_\post$ a distance $\sim 1/\bias{12}(q_i)$ from the edge of the
%window, while the pointer distribution takes essentially a
%Lorentzian shape of characteristic width $\sim \bias{12}(q_i)$,
%centered at $\wva{12}(q_\post)$.

In Figs.~\ref{fig:likeffex}c through ~\ref{fig:likeffex}e, we show
the likelihood effects on robust gaussian priors of variance
$\sigma_i^2$ at different locations. Here,  the prior sampling
region may be significantly altered while still preserving a
gaussian profile with relatively narrow width. For general initial
and final state, these effects can be described by performing a
gaussian approximation of the posterior distribution around its
maximum $q_\post$, determined by the equation
\begin{equation}\label{betaxep}
q_\post =  q_i  -2\sigma_i^2 {\rm Im} \frac{\bra{
\psf}{\hat{A}e^{i \hat{A} q_\post}}\ket{\pin}}{\bra{ \psf}{e^{i
\hat{A} q_\post}}\ket{\pin}}\, ,
\end{equation}
showing that the imaginary part of the complex weak value can be
interpreted as a ``bias function" for the posterior sampling
point.
%This approximation yields for the posterior gaussian
%distribution, centered at $q_\post$, and with a new variance
%\begin{equation}
%\sigma_\post^2 \simeq \frac{\sigma_i^2}{1 -\sigma_i^2
%\bias{12}'(q_\post)} \, .
%\end{equation}
%Note that the denominator  sets a necessary condition $\sigma_i^2
%\bias{12}'(q_\post) \lesssim 1$ for the validity of the gaussian
%approximation. If $\wva{12}(q)$ is  slowly varying within $\pm
%\sigma_\post$ of $q_\post$,
The resulting pointer distribution will be approximately a
gaussian centered at $p =\wva{12}(q_\post) $ with a corrected
width determined by the gaussian approximation. Two interesting
effects are then worth noting from these examples: First, as
illustrated in Fig. ~\ref{fig:likeffex}d, if the bias function is
large at the  prior sampling point $q_i$, the posterior sampling
point $q_\post$ may lie in the tail region of the prior
distribution. Thus, even if the prior distribution is quite
narrow, the sampled weak value $\wva{12}(q_\post)$ may differ
significantly from the  weak value $\wva{12}(q_i)$ at the prior
sampling point. The second effect has to do with appreciable
alterations of the widths as illustrated in Figs.
~\ref{fig:likeffex}c and ~\ref{fig:likeffex}e: if the prior
sampling point $q_i$ is set at a minimum (maximum) of the
likelihood function  (cases for which $q_\post = q_i$ in the
gaussian approximation), the respective posterior distributions in
$q$ will be widened (narrowed) with respect to the prior;
correspondingly, the pointer distributions may be narrowed
(widened) with respect to the prior
 distribution $\mathcal{P}(p|\phi)$. In particular, it follows that
 gaussian measurement  conditions probing the most ``eccentric" weak value
 on the orbit will generically show a \emph{squeeze} of the prior pointer
 distribution--a  surprising effect if the statistics are viewed
 as the result of sampling eigenvalues.

\begin{figure}
   \epsfxsize=3.45truein
\centerline{\epsffile{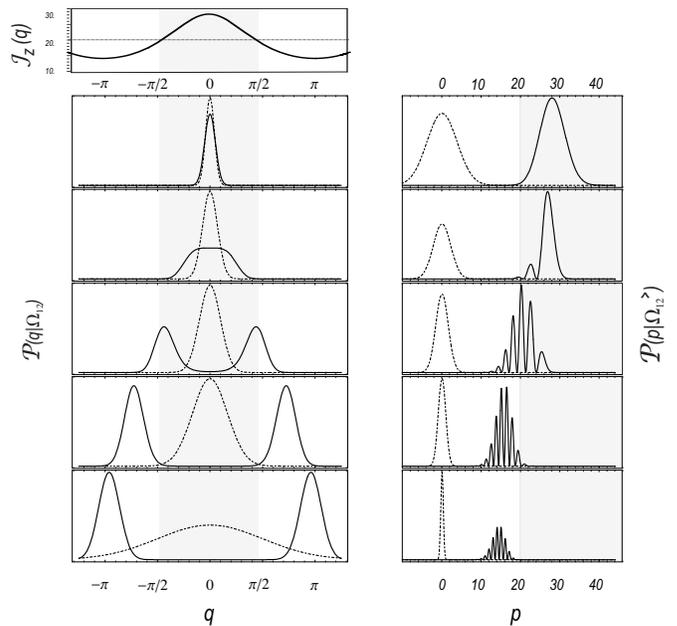}} \caption[\, ]{A weak to strong
measurement transition illustrated by priors of increasing width
in $q$ around $q=0$. Graphs follow the same convention as those of
Fig. 5.  }\label{fig:trans}
 \end{figure}

Turning finally to Figs.~\ref{fig:likeffex}f and
~\ref{fig:likeffex}g, we show the effects on two quite dissimilar
non-robust priors centered at $q=0$: a wide window function of
width $\varepsilon = 3 \pi$,  and a narrow Lorentzian of
half-width $\Gamma = \pi/24$ (comparable to the prior widths  in
Figs.~\ref{fig:likeffex}a and~\ref{fig:likeffex}c),  both
  encompassing the maximum likelihood regions around $q =
\pm \pi$ with either no suppression or insufficiently slow tail
suppression  of the likelihood factor. The resulting posterior
distributions in $q$ are then both qualitatively very similar and
similar in turn to the likelihood factor $\propto
\mathcal{P}_{12}(q)$ within the region $q \in [-3\pi/2,3\pi/2]$,
which from\rref{likeapp} is spproximately the sum of two
equally-shaped narrow gaussians at $q = \pm \pi$. Thus, conditions
are achieved for the superposition of two weak measurements at
$q_\post = \pm \pi$, both sampling in this case the least
eccentric weak value on the orbit, $\mathcal{J}_z(\pm \pi) =
j/\sqrt{2}$. The two peaks in these cases are in fact quite
similar to the single peak from the gaussian profile of
Fig.~\ref{fig:likeffex}e
 at $q =\pi$; thus, even while the prior pointer distributions
in~\ref{fig:likeffex}e through~\ref{fig:likeffex}f differ
substantially in their shapes, the resulting PPME pointer
distributions for all three cases share essentially the same
envelope, with the last two cases showing interference fringes
from  the superposition of the two weak measurement sampling
points.  This  interference pattern can then be connected to the
spectral distribution expected from a strong measurement: given
weak value and likelihood curves symmetric about $q=0$, and a
posterior distribution in $q$ with two similarly-shaped narrow
peaks at locations $q = \pm q_\post$, the resulting PPME pointer
distribution will  be the PPME pointer distibution for the single
peak weak measurement at $q_\post$, but modulated by the term
\begin{equation}
2\cos^2\left( 2 p q_\post - \delta(q_\post)\right) , \ \ \
\delta(q) = \int_{-q}^{q}dq'\, \wva{12}(q')\, ,
\end{equation}
 describing the interference pattern.
 For the situation depicted in Fig.~\ref{fig:likeffex}, the phase shift is
easily obtained from Eq.\rref{wvalj} and is given by
\begin{equation}
\delta(q) = 4 j \tan^{-1}\left(\sqrt{2}
\tan\left(\frac{q}{2}\right)\right)\, .
\end{equation}
For $q_\post \rightarrow \pi$, we have $\delta(q_\post)
\rightarrow 2 \pi j$;  hence,  interference patterns similar to
those of Figs~\ref{fig:likeffex}f and ~\ref{fig:likeffex}g will
show maxima at integer values of $p$ (corresponding to integer
values of $j$), or at half-integer values of $p$ for half-integer
$j$, consistently with spectrum of $\hat{J}_z$.

The foregoing  suggests a fairly general  picture underlying the
transition from weak to strong measurement conditions for  fixed
initial and final conditions, as the width of the prior
distribution in $q$ is varied. Illustrating this passage with a
gaussian prior of variable width $\sigma_i$  centered at $q=0$
(Fig.~\ref{fig:trans}) for the same $\hat{J}_z$ measurement, we
find the onset of a transitional behavior at  a critical value of
$\sigma_i$ (Fig.~\ref{fig:trans}b) where the gaussian
approximation fails. Beyond this critical value,  the exponential
rise of the likelihood factor dominates the prior on both sides,
thus producing two symmetrically opposed peaks, the locations of
which gradually move towards $q = \pm \pi$ as $\sigma_i$ is
increased. This transitional behavior is reflected in the
resulting pointer distribution by the emergence of an interference
pattern with increasingly closer fringes, modulated by an envelope
that gradually shifts with the sampled weak value
$\wva{12}(q_\post)$ from the eccentric to the normal region of
expectation. The pattern eventually settles at the characteristic
shape of the strong measurement distribution when the location of
the two peaks reaches $q = \pm \pi$,  only becoming sharper with
increasing $\sigma_i$  when the tails of the gaussian prior
``activate" the next likelihood peaks at $q = \pm 3\pi, \pm 5\pi$,
etc.

\section{Classical Correspondence of QAWV Framework.}
\label{Clas}

The connection between  macroscopic ``classical" properties and
weak values  has already been suggested in the
literature\ccite{AV90, Tanaka,Parks03}. In this section we give
further evidence of this connection by  showing the correspondence
of the QAWV framework in the classical limit. In particular, we
show that in the semi-classical limit, the necessary conditions
for a precise measurement of a classical dynamical quantity $A$
according to classical mechanics are at the same time the
conditions that in the quantum description guarantee a weak
measurement of the corresponding observable $\hat{A}$ yielding the
same numerical outcome.

 Let $x$ be the
configuration variable of a classical system, with free dynamics
described by the Lagrangian $L_o(\dot{x},x,t)$. For simplicity, we
concentrate on a measurement of a function $A(x)$ of the
configuration variable $x$ alone,  with a measurement Lagrangian
of the form
\begin{equation}
L_M(q,x,t) =  \delta(t-t_i) A(x) q \, ,
\end{equation}
coupling the system and an external classical apparatus with
pointer variable $p$ and canonical conjugate  $q$. To connect with
the results of section \ref{qaves}, we interpret the  pre- and
post- selection  as the fixing of initial and  final boundary
conditions on the system trajectory: $x_1 \equiv x(t_1)$ and $x_2
\equiv x(t_2)$ with $t_2 > t_i > t_1$.

Let us assume for simplicity throughout that only one solution is
possible for the Euler-Lagrange equations. For non-zero $q$, the
trajectory of the system will differ from its free trajectory due
to a modification of the equations of motion by an additional
$q$-dependent impulsive force $ F_M=\delta(t-t_i) A'(x) q \, $
arising from the back reaction of the apparatus on the system.
Then, since the actual trajectory will be some function
$x_{12}(t;q) = x(t;x_1,x_2,q)$ of the boundary conditions and $q$,
the quantity $A(x(t_i))$  will generally depend on $q$ as well. In
analogy with our previous notation, define the function
\begin{equation}
\widetilde{\mathcal{A}}_{12}(q) \equiv A(x_{12}(t;q)) \, .
\end{equation}
As one can show from the equations of motion,
 the classical
action for the total Lagrangian $L_T = L_o +L_M$ evaluated on the
trajectory $x_{12}(t;q)$,
\begin{equation}\label{clasact}
\widetilde{S}(q|x_1 x_2) \equiv \int_{t_1}^{t_2} dt\,
L_T[{x_{12}(t;q)}] \, ,
\end{equation}
serves as a generating function for
$\widetilde{\mathcal{A}}_{12}(q)$, i.e.,
$\widetilde{\mathcal{A}}_{12}(q) =\widetilde{S}_{12}'(q)$. Thus,
from the equations of motion for the apparatus, we find that the
pointer variable suffers the impulse $p$ at the time $t_i$
\begin{equation}
p_{\aft} = p_{\bef} +  \widetilde{\mathcal{A}}_{12}(q) = p_{\bef}
+  \widetilde{S}_{12}'(q)\, ,
\end{equation}
in direct correspondence with Eq.\rref{canshift}.

We now turn  to the probabilistic aspects of the measurement.
Allowing for uncertainties in the initial state (i.e., the point
in phase space) of the apparatus,  we describe our knowledge with
a prior p.d.f. ${\cal P}(qp|I^\bef)$ for the state of the
apparatus before the measurement, where $I$ denotes all available
prior information. We also assume that initial conditions on the
system are irrelevant for this prior assessment of probabilities
so that ${\cal P}(qp|I x_1^\bef) ={\cal P}(qp|I ^\bef)$. Since the
variable $q$ enters the equations of motion of the system,
knowledge of the final condition $x_2$ becomes relevant for
inferences about $q$ at the time of the measuring interaction, and
will therefore determine a re-assessment of prior probabilities.
We must therefore compute the posterior p.d.f. ${\cal P}(qp|I x_1
x_2 ^\bef)$ for the apparatus, conditioned on the endpoints of the
system trajectory, at the time \textit{before} the interaction.
The dynamics of the measurement can then be described by the
Liouville evolution generated by $\widetilde{S}_{12}(q)$, i.e.,
\begin{equation}\label{classtatetarns}
{\cal P}(qp|I x_1 x_2 ^\aft) =
e^{-\widetilde{\mathcal{A}}_{12}(q)\frac{\partial}{\partial
p}}{\cal P}(qp|I x_1 x_2 ^\bef) \, .
\end{equation}
Using Bayes' theorem, we find that
\begin{equation}\label{clasBayes}
{\cal P}(qp|I x_1 x_2 ^\bef) = \frac{{\cal P}(x_2| x_1 q )}{{\cal
P}(x_2|I x_1 )} {\cal P}(qp|I^\bef )\, ,
\end{equation}
where  we have used the fact that  $q$ is the only relevant
apparatus variable entering the dynamics of the system, thus
yielding a likelihood factor ${\cal P}(x_2| x_1 q )$ analogous to
$P_{12}(q)$ in the quantum case. Finally, evolving to the time
after the measurement through Eq.\rref{classtatetarns} and
marginalizing, we obtain for the pointer variable distribution
after the measurement:
\begin{equation}\label{claspdf}
{\cal P}(p|I x_1 x_2^\aft) =\left \langle\, \delta\!\left(\, p -
p' - g \widetilde{\mathcal{A}}_{12}(q')\, \right) \,\right
\rangle_\bef
%\int\! dp'\! dq'\, \delta(p -
%p' - g \widetilde{\mathcal{A}}_{12}(q') ) {\cal P}_{-\epsilon}(q'p'|I
%x_1 x_2) \, ,
\end{equation}
where the dummy variables $q"$ and $p'$ are averaged over the
reassessed initial phase space p.d.f. for the apparatus ${\cal
P}(q'p'|I x_1 x_2^\bef)$. This distribution is in complete analogy
with Eq.\rref{quantpdf}
 if averages over ${\cal P}(qp|I x_1 x_2^\bef) $
are identified with averages over the reassessed state
$\ket{\tilde{\phi}_{_{12}}^\bef}$ and if
$\widetilde{\mathcal{A}}_{12}(q)$ is identified
 with the $q$-dependent weak value $\wva{12}(q)$. With this identification,
 Eq. \rref{finalmoms} for the associated moments can be used
 for both the classical or quantum
  descriptions. Furthermore, the terms
  $\langle p \rangle_\bef$ and $\langle
\{\Delta p ,\Delta \mathcal{A} \} \rangle_\bef$ in
\rref{finalmoms} can also be eliminated in the classical case by
requiring that the prior phase space distribution factors as
${\cal P}(qp|I^\bef)={\cal P}(q|I^\bef){\cal P}(p|I^\bef)$ with
the expectation value of $p$ vanishing over ${\cal P}(p|I^\bef)$.

We  can now show that under appropriate semi-classical conditions
on a corresponding quantum system, the above analogy is not only
formal but rather constitutes a true numerical correspondence
between  classical and quantum averages. For this, we need to
calculate the so-far unspecified likelihood factor ${\cal P}(x_2|
x_1 q )$ in Eq.\rref{clasBayes}, which plays the role of
 ${\cal P}_{12}(q)$ in the state
reassessment of Eq.\rref{statereassess}. In the classical
description, the
% trajectory of the system is determined from
%initial boundary conditions when both the configuration variable
%$x$ and its conjugate momentum (denote it by $\pi$) are known. In
%such a case, the trajectory (in the presence of the measurement)
%will be some function $x(t;x, \pi, q, t_1)$. However, the
%specification of only  the configuration variable $x=x_1$ leaves
%the momentum $\pi$ completely unspecified, so that the
probability of being at $x_2$ at the time $t_2$ is proportional to
the integral $\int  d\pi \delta(x_2 -x(t_2;x_1, \pi, q, t_1))$
over all possible initial momenta $\pi$ of the system, yielding
\begin{equation}
{\cal P}(x_2|x_1 q)   \propto  \left|\frac{\partial\pi_1
}{\partial x_2} \right|
 \, ,
\end{equation}
where $\pi_1 = \pi(t_1;x_1,x_2,q)$ is the value of the initial
momentum as determined from the boundary conditions. This initial
momentum can  be obtained from a variation of the classical
action, $\pi_1 = -\partial_{x_1}
\widetilde{S}_{12}(q)$\ccite{Arnold}, so that
\begin{equation}
{\cal P}(x_2|x_1 q) \propto \left|\frac{\partial^2
\widetilde{S}_{12}(q) }{\partial x_1 \partial x_2}\right|\, ,
\end{equation}
(known as  Van Vleck determinant\ccite{Cecile} from its extension
to higher dimensions). Correspondence with the quantum description
can now be established by calculating the quantum mechanical
propagator $ \langle x_2|\hat{U}(t_2,t_1;q)| x_1 \rangle $ for the
corresponding quantum system, with $\hat{U}(t_2,t_1;q)$ being the
time evolution operator associated with the classical Lagrangian
$L_o + L_M(q)$. As  is easily verified, this   is the relevant
amplitude for the von Neumann measurement of $A(\hat{x})$ at the
time $t_i$ with the given boundary conditions. Under appropriate
semiclassical conditions\ccite{Cecile} (e.g., small times, large
masses, slowly varying potentials, etc.), the propagator reduces
to the semiclassical or WKB form
\begin{equation}
\langle x_2|\hat{U}(t_2,t_1;q)| x_1 \rangle
\stackrel{WKB}{\longrightarrow}\frac{1}{(2 \pi i)^{\frac{1}{2}}}
\sqrt{\left|\frac{\partial^2 \widetilde{S}_{12}(q) }{\partial x_1
\partial x_2}\right|}\ e^{i \widetilde{S}_{12}(q)} \, ,
\end{equation}
where $\widetilde{S}_{12}$ is  classical action of
Eq.\rref{clasact}. Consequently, under semiclassical conditions,
the weak value $\wva{12}(q)$ of $A(\hat{x})$ at the time $t_i$
 coincides with the classical $\widetilde{\mathcal{A}}_{12}(q)$; similarly, the
likelihood factor in the re-assessment of the initial state of the
apparatus (Eq.~\ref{statereassess}) is  the square root of the the
likelihood factor $\propto \left|\partial_1
\partial_2 \widetilde{S}_{12}(q) \right|$  involved in the re-assessment  probabilities in the classical
description.

Thus, assuming the conditions ensuring $\langle p \rangle_\bef$=0,
the final posterior mean value of $p$ will be given both in the
classical and quantum descriptions by the average value $\langle
\mathcal{A}\rangle_\bef$ over the respective posterior
distributions in $q$, which can be made to coincide. This allows
us to claim
  a stronger correspondence between the classical
and quantum  descriptions when the system satisfies semiclassical
conditions: for the same prior distributions in $q$,
  the classical and quantum
expectation values and variances of $A$ \emph{are numerically
equal} and hence, in particular,   the  final pointer expectation
values are equal. It follows that the minimum dispersion
conditions on the variable $q$  that  in a classical description
are required for a precise measurement of $A$ (i.e., $\Delta q
\rightarrow 0 \Rightarrow \Delta A \rightarrow 0$), are at the
same time the conditions that in the quantum description will
guarantee a \emph{weak} measurement of $\hat{A}$ yielding the same
numerical value. This correspondence strongly suggests that
indeed, what we call macroscopic ``classical" properties, are in
fact weak values.

Let us  elaborate on this assertion: The use of classical
mechanics to describe macroscopic systems
 or other quantum systems exhibiting classical behavior
 relies on the  fact that  individual measurements  may be devised
 so that:  a) the effect on the measurement device accurately reflects
the numerical value of the classical observable being measured,
  b) no appreciable disturbance is
produced on the system as a result of the measurement interaction;
and c) the effect on the measurement device is statistically
distinguishable (i.e., the signal to noise ratio is large). The
three conditions can be stated as follows: a) $\frac{\Delta
\mathcal{A}}{\mathcal{A}} \ll 1$, b) $  \ave{q}=0$, $\Delta q
\rightarrow 0$ and c) $\frac{\Delta p }{\mathcal{A}} \gg 1$. In
the quantum description, conditions  a) and b) are weak
measurement conditions and can be attained asymptotically by
making the posterior uncertainty $ \Delta q_\aft$ tend to zero,
with the posterior average fixed at $q=0$; however, condition c)
cannot be upheld in the limit $ \Delta q \rightarrow 0$ since
$\Delta p \rightarrow \infty$ due to the uncertainty principle.
Equivalently, conditions a) and b) cannot be fulfilled if
condition c) is to be satisfied by demanding $\Delta p \rightarrow
0$ as in the case of an ideally strong measurement.

While it is therefore impossible to satisfy the three conditions
either in the absolute strong or weak limits, relatively weak
measurement conditions can nevertheless be found as a compromise
in the uncertainty relations so that conditions a), b) and c) are
simultaneously satisfied ``for all practical purposes" when
classical-like physical quantities are involved.  Indeed, for such
quantities one expects $\mathcal{A}$ to be in a sense ``large"
relative to atomic scales, or more precisely, to scale extensively
with some scale parameter $\lambda$ growing with the size or
``classicality" of the system (such the mass, or the number of
atoms).  One can then choose a scaling relation for $\Delta q$,
i.e., $\Delta q \sim \lambda^{-\gamma}$ so that
\begin{equation}
\frac{\Delta p}{ \mathcal{A} } \sim \frac{\Delta \mathcal{A}
}{\mathcal{A} } \ll 1 \, ,
\end{equation}
in which case  conditions a) b) and c) can be satisfied in the
limit $\lambda \rightarrow \infty$. Assuming that $\mathcal{A}'$
scales as $\mathcal{A}$, then with the aid of the uncertainty
relation $\Delta p \sim 1/\Delta q$ and $\Delta \mathcal{A} \simeq
\mathcal{A}' \Delta q$, we find that this is possible in the
quantum description if $\Delta q$ can be made to scale as
\begin{equation}
\Delta q \sim \lambda^{-1/2} \, ,
\end{equation}
in which case
\begin{equation}
\frac{\Delta p }{\mathcal{A}} \sim  \frac{\Delta
\mathcal{A}}{\mathcal{A}} \sim \lambda^{-\frac{1}{2}} \, .
\end{equation}
As was recently  shown\ccite{Poullin}, this is precisely the
 scaling relation of the  optimal compromise for measurements of
 ``classical" collective properties (such as center of mass position or total momentum) of
a large number $(\sim \lambda)$ of independent atomic
constituents.

\section{Conclusion}
\label{concl}

In this paper we have  advanced the claim that
 weak values of  quantum mechanical observables constitute legitimate
physical concepts providing  an objective description of the
properties of a quantum system known to belong to a completely
pre- and postselected ensemble. This we have done by addressing
two aspects, namely the physical interpretation of weak values,
and their applicability as a physical concept outside the weak
measurement context.

Regarding the physical meaning of weak values, we have shown  that
the weak value corresponds to a definite mechanical response of an
ideal measuring probe the effect of which, from the point of the
system, can be described as an infinitesimally uncertain unitary
transformation. We have  stressed  how from this operational
definition  the weak value of an observable $\hat{A}$ is tied to
the role of $\hat{A}$ as a generator of infinitesimal unitary
transformations.  We believe that this sharper operational
formulation of weak
 values in terms of  well-defined mechanical effects
clarifies the sense in which  weak values describe
 new and surprising  features of the quantum domain.
Regarding the applicability of the concept of weak values in more
general contexts, we have shown that arbitrary-strength von
Neumann measurements can be analyzed in the framework of quantum
averages of weak values, in which dispersion in the apparatus
variable driving the back-reaction on the system entails a quantum
sampling of  weak values. The framework has been shown to merge
naturally into the classical inferential framework in the
semi-classical limit.

 It is our hope that the framework introduced in the present
paper may serve as a motivation for a refreshed analysis of the
measurement process in quantum mechanics.

\section{Acknowledgments}

Y. A. acknowledges support from the Basic Research Foundation of
the Israeli Academy of Sciences and Humanities and the National
Science Foundation. A.B. acknowledges support
  from Colciencias (contract No. 245-2003). This paper is based in
  part on the latter's doctoral dissertation\ccite{BoteroThesis},
 the completion of which  owes much to  Prof. Yuval Ne'eman and financial
    support from Colciencias-BID II and a one-year scholarship from
  ICSC - World Laboratory.

\end{document}